\documentclass[12pt]{article}

\usepackage{graphicx}%
\usepackage{dcolumn}%
\usepackage{bm}
\usepackage{tabularx}
\usepackage{appendix}
\usepackage{graphicx}
\usepackage{subcaption}
\usepackage{colortbl}
\usepackage{color}
\usepackage{cancel}
\usepackage{comment}
\usepackage[countmax]{subfloat}
\usepackage{framed}
\usepackage{mdframed}
\usepackage{multirow}
\usepackage{slashed}
\usepackage{booktabs} 
\usepackage{hyperref}
\usepackage{caption}
\usepackage{ragged2e}
\DeclareCaptionJustification{justified}{\justifying}
\usepackage{physics}
\usepackage{comment}
\usepackage{mathrsfs}
\usepackage{ulem}
\usepackage{amssymb}

\begin{document}

\begin{titlepage}
\begin{flushright}
{KOBE-COSMO-25-22}
\end{flushright}

\vspace{50pt}

\begin{center}

{\large{\textbf{Chiral gravitational waves from domain walls\\ in Nieh-Yan gravity }}}

\vspace{55pt}
\shortstack[l]
{Jiro Soda$^{\flat}$ and Maki Takeuchi$^{*}$}

\vspace{20pt}
\shortstack[l]
{\it{\small $^\flat$ Department of Physics, Kobe University, Kobe 657-8501, Japan}}

{\it {\small $^*$Graduate School of Sciences and Technology for Innovation, }} \\
\noindent
{\it {\small \ \  Yamaguchi University, Yamaguchi-shi, Yamaguchi 753-8512, Japan} }
\vspace{58pt}
\end{center}
\begin{abstract}
We study the  scattering of gravitational waves by axion domain walls in teleparallel gravity with the Nieh-Yan term. 
Since a domain wall causes the parity violation, the transmitted 
gravitational waves also exhibit the parity violation.
We calculate the degree of circular polarization of gravitational waves. It turns out that gravitational waves after going through the domain wall could be chiral.
Remarkably, the degree of circular polarization does not depend on the tension of the domain wall.
\end{abstract}

\end{titlepage}

\tableofcontents

\section{Introduction}
\label{sec:introduction}

Parity violation is ubiquitous in nature.
In fact, parity violation has played an important role not only in particle physics but also in cosmology.
For example, recently, the violation of parity in  galaxy distributions has been discussed~
\cite{Philcox:2022hkh,Creque-Sarbinowski:2023wmb,Coulton:2023oug}.
Thus, it is worth exploring the parity violation in gravity.
In Einstein gravity, parity violation of spacetime is 
realized by the Chern-Simons term~\cite{Campbell:1990fu,Jackiw:2003pm,Lue:1998mq,Alexander:2009tp}. Interestingly, the Chern-Simons term induces parity violating gravitational waves
~\cite{Satoh:2007gn,Satoh:2008ck} in the early universe.
Moreover, when it is coupled with axion dark matter, the birefringence phenomena can be expected~\cite{Yoshida:2017cjl}.
However, it is known that there exist ghost modes in the spectrum~\cite{Dyda:2012rj}. Remarkably, we can avoid the appearance of ghosts in gravitational wave modes in Horava gravity where general coordinate invariance is violated~\cite{Takahashi:2009wc}.
Hence, the question is whether there exists a general coordinate invariant parity violation in gravity without ghost instability.
Interestingly, in the context of teleparalell gravity,
the Nieh-Yan term can violate parity symmetry~\cite{Nieh:1981ww,Nieh:2008btw}. Remarkably, it has been shown that the Nieh-Yan term does not have ghost instability~\cite{Li:2020xjt,Li:2021wij}. It should be noted that, within the framework of Einstein-Cartan gravity~\cite{BeltranJimenez:2019esp}, it is possible to have various terms that violate parity.
The Nieh-Yan term has also been studied in cosmology~\cite{Wu:2021ndf,Cai:2021uup,Xu:2024kwy,Adshead:2025xtq}.

In most gravitational theories, what induces the parity violation is an axion. The axions were originally introduced into QCD to resolve the strong CP problem~\cite{Peccei:1977hh,Weinberg:1977ma,Wilczek:1977pj,Kim:1979if,Shifman:1979if,Dine:1981rt,Zhitnitsky:1980tq}. Recently, axion like particles have been intensively studied in the cosmological context~\cite{Marsh:2015xka}. In fact, axion-like particles are ubiquitous in string theory~\cite{Svrcek:2006yi}.
 Intriguingly, when the Peccei-Quinn (PQ) symmetry is spontaneously broken after inflation, domain walls form~\cite{Sikivie:1982qv}. The formation of axion domain walls in the early universe is an important phenomenon~\cite{vilenkin}. The tension of the domain walls formed in such scenarios is of the order of $ (10^{9}\sim 10^{12} {\rm GeV})^3$~\cite{Hiramatsu:2012gg,Graham:2015ouw}. However, since their energy density decays too slowly relative to the energy density of the surrounding matter, they overclose the universe~\cite{Zeldovich:1974uw}. 
 This is a notorious domain wall problem.
 In addition to the domain wall problem, we have to take into account observational constraints. In fact, CMB observations give a stringent constraint on the tension of domain walls $\sigma < (0.93\ {\rm{MeV}})^3 $ at the 95\% confidence level for standard $\Lambda$-CDM cosmology~\cite{Lazanu:2015fua}.

There are several ideas to solve the domain wall problem. One can simply consider the pre-inflationary breakdown of the PQ symmetry. In~\cite{Larsson:1996sp}, the authors consider a tilt of the potential or biased initial conditions.
Interestingly, in the scenario \cite{Babichev:2021uvl}, the domain wall does not have the domain wall problem because the energy density of the domain wall decreases faster than that of radiation. In any scenario, domain walls do exist
in the history of the universe. 

 Since the axion induces both domain walls and parity violation in gravity,
 it is interesting to study interaction between gravitational waves
 and domain walls. 
 In the case of Chern-Simons gravity, the scattering of gravitational waves by the domain wall has been studied~\cite{Kanno:2023kdi}.
 However, as we already mentioned, ghosts appear in this case.
In this paper, we study the scattering of gravitational waves by axion domain walls in Nieh-Yan gravity, where ghosts do not appear. We can neglect the cosmic expansion because the relevant scattering typically occurs well within the Hubble horizon. Since axions can couple with gravitational waves through the Nieh-Yan term, gravitational waves show birefringence. 
We will clarify to what extent the parity invariance is violated in the presence of domain walls. More precisely, we calculate the degree of circular polarization of gravitational waves. We will see that chiral gravitational waves in Nieh-Yan gravity are ubiquitous in the presence of domain walls.

The organization of the paper is as follows. In Section 2, we review teleparallel gravity. In section 3, equations of gravitational waves in the presence of domain wall are derived.
In section 4, we study the scattering of gravitational waves by a domain wall
and evaluate the transmission probability of polarized gravitational waves. Then, we evaluate the degree of circular polarization of the gravitational waves. The final section is devoted to the conclusion.

\section{Teleparallel gravity with the Nieh-Yan term}
In this section, we start with a brief review of tereparallel gravity (TG).
The TG theory is constructed by the tetrad fields $e^{A}_{\mu}$ which were used to construct the spacetime metric $g_{\mu \nu}=\eta_{AB}e^{A}_{\mu}e^{B}_{\nu}$. Here $\eta_{AB}={\rm{diag}}\{-1,+1,+1,+1\}$ is the local Minkowskian spacetime metric.
We use the tetrad indices by $A,B,\cdots=0,1,2,3$ and $a,b,\cdots=1,2,3$. The spacetime tensor indices are expressed by $\mu,\nu,\cdots=0,1,2,3$ and $i,j,\cdots=1,2,3$. 
The antisymmetric symbol $\epsilon^{\mu \nu \rho \sigma}$ satisfies $\epsilon^{0i j k}=-\epsilon^{i j k}$.

Let us define a covariant derivative
\begin{eqnarray}
   \hat{\nabla}_\mu A^\nu \equiv \partial_\mu A^\nu +\hat{\Gamma}_\mu {}^\nu{}_\lambda A^\lambda
\end{eqnarray}
The TG theory is identified as 
\begin{align}
    \hat{R}^\rho{}_{\mu\nu\lambda}&=0, \label{flat} \\
    \hat{\nabla}_{\alpha}g_{\mu \nu}&=0, \label{metricity}\\
    T^\rho{}_{\mu\nu}&\neq0, 
\end{align}
where $\hat{R}^\rho{}_{\mu\nu\lambda}$ and $T^\rho{}_{\mu\nu}$ denote the Riemann tensor and the torsion tensor, respectively. The Riemann tensor vanishes 
\begin{eqnarray}
    \hat{R}^\rho{}_{\mu\nu\lambda}  = \partial_\nu\hat{\Gamma}^\rho{}_{\mu\lambda} - \partial_\lambda\hat{\Gamma}^\rho{}_{\mu\nu}
    + \hat{\Gamma}^\rho{}_{\alpha\nu} \hat{\Gamma}^\alpha{}_{\mu\lambda}
    - \hat{\Gamma}^\rho{}_{\alpha\lambda} \hat{\Gamma}^\alpha{}_{\mu\nu}=0,
\end{eqnarray}
where $\hat{\Gamma}^\rho{}_{\mu\nu}$ denotes the terepararell connection.
The torsion is defined by 
\begin{eqnarray}
    T^\rho{}_{\mu\nu} = \hat{\Gamma}^\rho{}_{\mu\nu} - \hat{\Gamma}^\rho{}_{\nu\mu}.
\end{eqnarray}
From the metricity (\ref{metricity}), we get the following
\begin{eqnarray}
    \hat{\Gamma}^\rho{}_{\mu\nu} =  \Gamma^{\rho}{}_{\mu\nu} + K^\rho{}_{\mu\nu},
    \label{crirelation}
\end{eqnarray}
where the Cristoffel symbol $\Gamma^{\rho}{}_{\mu\nu}$ is defined by
\begin{eqnarray}
    \Gamma^{\rho}{}_{\mu\nu} 
    = \frac{1}{2} g^{\rho\lambda}
    \left( g_{\lambda\mu,\nu} + g_{\lambda\nu,\mu} - g_{\mu\nu,\lambda}\right),
\end{eqnarray}
and the contortion $K^\rho{}_{\nu\mu}$ is given by
\begin{eqnarray}
    K^\rho{}_{\nu\mu}  =\frac{1}{2} T^\rho{}_{\nu\mu} + T_{(\nu}{}^{\rho}{}_{\mu)} \ .   \label{contorsion}
\end{eqnarray}
From Eq.\eqref{crirelation}, the Riemann tensor is expressed as
\begin{eqnarray}
    \hat{R}^\rho{}_{\mu\nu\lambda}  = 
     R^\rho{}_{\mu\nu\lambda} + \nabla_{\nu} K^\rho{}_{\lambda\mu}  
     -\nabla_{\lambda} K^\rho{}_{\nu\mu}
+K^\rho{}_{\nu\sigma}K^\sigma{}_{\lambda\mu}
-K^\rho{}_{\lambda\sigma}K^\sigma{}_{\nu\mu} \ ,
\end{eqnarray}
where the tensor $ R^\rho{}_{\mu\nu\lambda}$
is constructed by the Cristoffel symbol.
From the flat condition (\ref{flat}), we have a relation
\begin{eqnarray}
 R^\rho{}_{\mu\nu\lambda} &=- \nabla_{\nu} K^\rho{}_{\lambda\mu}  
     +\nabla_{\lambda} K^\rho{}_{\nu\mu}
-K^\rho{}_{\nu\sigma}K^\sigma{}_{\lambda\mu}
+K^\rho{}_{\lambda\sigma}K^\sigma{}_{\nu\mu}, 
\end{eqnarray}
which gives rise to a relation
\begin{align}
R = \nabla_{\rho}(- K^\rho{}_{\mu}{}^{\mu}  
     +K^\mu{}_{\mu}{}^{\rho})
-K^\rho{}_{\rho\sigma}K^{\sigma \mu}{}{}_{\mu}
+K^\rho{}_{\mu\sigma}K^\sigma{}_{\rho}{}^{\mu} \ .
\label{Rcal}
\end{align}
The definition of the contorsion leads to the following
\begin{align}
  -K^\rho{}_{\rho\sigma}K^{\sigma \mu}{}{}_{\mu}
+K^\rho{}_{\mu\sigma}K^\sigma{}_{\rho}{}^{\mu}=  T^\rho{}_{\rho\mu}T^\sigma{}_{\sigma}{}^{\mu}-\frac{1}{4}T^{\rho \sigma \mu}T_{\rho \sigma \mu}-\frac{1}{2}T^{\mu \sigma \rho}T_{\rho \sigma \mu}   \ .
\end{align}
The action for teleparallell gravity is given by
\begin{eqnarray}
  S= \frac{M_{\rm{p}}^2}{2} \int d^4x \sqrt{-g} \left[   
 T^\rho{}_{\rho\mu}T^\sigma{}_{\sigma}{}^{\mu}-\frac{1}{4}T^{\rho \sigma \mu}T_{\rho \sigma \mu}-\frac{1}{2}T^{\mu \sigma \rho}T_{\rho \sigma \mu}\right] \ , \label{Tel}
\end{eqnarray}
where $M_{\rm{p}} $ is the reduced Planck mass.
From the relation (\ref{Rcal}), we see that the above action is equivalent to Einstein gravity at this level.

Let us introduce an axion-like field $\phi$. Now, we can add
the Nieh-Yan term coupled to an axion-like field to the TG gravity. It takes the following form
\begin{align}
    S_{\rm{NY}}=\int d^4 x \sqrt{-g}\frac{M_{\rm{p}}^2\ell  }{4} \phi T_{A \mu \nu}\tilde{T}^{A \mu \nu}  \ ,
\end{align}
where  $\ell$ is a coupling constant representing the charcteristic length scale and $\tilde{T}^{A \mu \nu}=\frac{1}{2}\varepsilon^{\mu \nu \rho \sigma} T^A{}_{\rho \sigma}$ is a dual of the torsion tensor. Here, the Levi-Civita tensor $\varepsilon^{\mu \nu \rho \sigma}$ is defined by the antisymmetric symbol $\epsilon^{\mu \nu \rho \sigma}$ as $\varepsilon^{0123}=\epsilon^{0123}/{\sqrt{-g}}=-1/{\sqrt{-g}}$.
In the language of differential forms, the torsion two-form is expressed as $T^A=d\omega^A + \omega^{A}{}_B \wedge \omega^B $.
Here, $\omega^B $ and $\omega^{A}{}_B$  are tetrad 1-form and
connection 1-form, respectively.
In teleparallel gravity, we can take the gauge $\omega^A{}_B =0$.
Thus, the torsion two form is expressed by
\begin{eqnarray}
    T^A{}_{\mu\nu} = \partial_\mu e^A_\nu -  \partial_\nu e^A_\mu,
\end{eqnarray}
Hence, the Nieh-Yan term is a total derivative in the absence of 
the axion field. 
Taking into account the equivalence between the Einstein-Hilbert action and the action (\ref{Tel}), the action we consider is the following
\begin{eqnarray}
   S = \int d^4x \sqrt{-g} \left[\frac{M_{\rm{p}}^2}{2} R 
   + \frac{M_{\rm{p}}^2\ell }{4} \phi T_{A \mu \nu}\tilde{T}^{A \mu \nu}\right].
   \label{action}
\end{eqnarray}
The fundamental variable is the tetrad $e^A_\mu$. In order to make the system consistent, we need to introduce the kinetic term and potential for the axion as we will do in the next section.

\section{Gravitatinal waves in domain wall background}

In the previous section, we have explained teleparallel gravity with the Nieh-Yan term. To make the Nieh-Yan term non-trivial, we
 incorporate the axion which is parity odd. Hence, if the axion has an expectation value, the parity can be violated. 
Domain walls are known to be ubiquitous in the presence of the axion. We shall derive equations for gravitational waves in the domain wall background. 

The action for the Nieh-Yan modified TG is given by
\begin{eqnarray}
   S = \int d^4x \sqrt{-g} \left\{\frac{M_{\rm{p}}^2}{2} R 
   + \frac{M_{\rm{p}}^2 \ell }{4}\phi T_{A \mu \nu}\tilde{T}^{A \mu \nu}-\left(\frac{1}{2}\partial_{\mu}\phi\partial^{\mu}\phi +V(\phi)\right) \right\}.
   \label{action}
\end{eqnarray}
Taking the variation with respect to $\phi$, we get
\begin{align}
     \square\, \phi-V_{\phi}=-\frac{M_{\rm{p}}^2 \ell}{4}T_{A \mu \nu}\tilde{T}^{A \mu \nu}.
    \label{EOM_phi}
\end{align}
The equation of motion for the tetrad field is
\begin{align}
    M_{\rm{p}}^2G^{\mu \nu}=T^{\mu \nu}-\frac{M_{\rm{p}}^2 \ell}{2}\partial_{\lambda}\phi \varepsilon^{\lambda \mu \alpha \beta}
    T^\nu{}_{\alpha \beta}.
\end{align}
The potential $V(\phi)$ is a double-well potential with two minima located at $\phi=\pm\eta$ such as
\begin{eqnarray}
    V (\phi )=\frac{1}{4} \lambda \left( \phi^2 -\eta^2 \right)^2,
\end{eqnarray}
where $\lambda$ is a coupling constant. We consider Minkowski spacetime and the domain wall to be static and planar; thus, without loss of generality, we place it in the $(x,y)$-plane, making it perpendicular to the $z$-axis. Under this setup, the corresponding domain-wall solution becomes
\begin{eqnarray}
   \phi (z)=\eta \tanh \sqrt{\frac{\lambda}{2}}\eta z .
\end{eqnarray}
Using this solution, the tension is calculated as
\begin{eqnarray}
    \label{sigma}
	\sigma=\int dz\,\phi^{\prime\,2}(z) \sim \sqrt{\lambda}\,\eta^3\ ,
\end{eqnarray}
where $\phi^{\prime}$ denotes the derivative with respect to $z$.

Given the above background, 
let us consider gravitational waves (GWs), represented by tensor-mode perturbations of the three-dimensional spatial metric:
\begin{align}
    ds^2=-dt^2+(\delta_{ij}+h_{ij})dx^i dx^j,
\end{align}
where $\delta_{ij}$ denotes Kronecker delta.
We take the transverse and traceless gauge $h_{ij,j}=h_{ii}=0$.
Here we consider only the linear perturbation terms as
\begin{eqnarray}
   - \ddot{h}^{ij} + \nabla^2 h^{ij} = - \ell \partial_k \phi \epsilon^{km(i} \dot{h}^{j)}{}_m,
   \label{EOM_h}
\end{eqnarray}
where we used the relation $e^A{}_{\mu}=\delta^A{}_{\mu}+\frac{1}{2}h^A{}_{\mu}$
and symetrized with respect to $i,j$.
The equation for the axion is decoupled in linear order. Hence, we do not consider this hereafter.

In this work, we investigate the scattering of GWs by the axion domain wall. For simplicity, we consider incident gravitational waves perpendicular to the domain wall. Even in generic cases, the qualitative result is similar. Since the domain wall violates the parity symmetry, GWs show birefringence. We evaluate the degree of the circular polarization after the transmission of GWs through the dimain wall.

\begin{figure}[t]
        \centering
        \includegraphics[width=12cm]{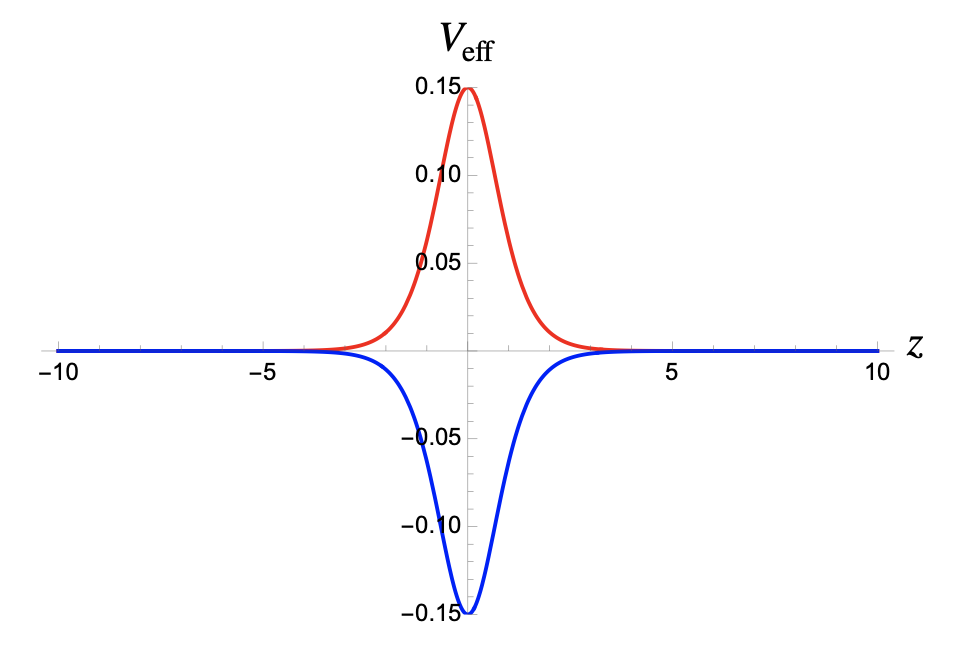}
        \vspace{-6pt}
        \caption{The plot of the effective potential $V_{\rm{eff}}$. The red and blue curves correspond to the right- and left-handed circular polarization modes, respectively. The parameters are set to be $\omega \ell =0.15$, $\eta=1$, and $\lambda=2$.}
        \label{fig_potential}
        \vspace{0.7cm}
\end{figure}

Since we assumed that GWs propagate along the $z$-axis, the wave vector is expressed as $k^{\mu}=(\omega,0,0,\omega)$ in the asymptotic region. We introduce the polarization tensors for the right-handed and left-handed circularly polarized modes as
%
	\begin{eqnarray}
	e^{(R)}_{ij}=
	\begin{pmatrix} 
		  1 & i & 0 \\
		  i & -1 & 0  \\
		  0 & 0 & 0  \\
	\end{pmatrix},
	\qquad
	e^{(L)}_{ij}=
	\begin{pmatrix} 
		  1 & -i & 0  \\
		  -i & -1 & 0  \\
		  0 & 0 & 0 \\
	\end{pmatrix}.
	\end{eqnarray}
Note that we have relations
\begin{eqnarray}
    \varepsilon^{zpk} e^{(R)}_{ik} = i e^{(R)}_{ip}\ ,
    \qquad
    \varepsilon^{zpk} e^{(L)}_{ik} = -i e^{(L)}_{ip}\ .
\end{eqnarray}
Using these polarization tensors, the GWs can be decomposed into the amplitude and the polarization such as
%
	\begin{eqnarray}
	\label{eq:hij}
	h_{ij}(t,z) = h_R(t,z) e^{(R)}_{ij} + h_L(t,z) e^{(L)}_{ij}\ .
    \label{h_RL}
	\end{eqnarray}
Here, $h_{R/L}$ denotes the amplitude of the right-handed/left-handed mode.
By substituting Eq.\eqref{h_RL} into Eq.\eqref{EOM_h}, we obtain
\begin{eqnarray}
   - \ddot{h}_{R/L}(t,z) + \frac{d^2 }{dz^2} h_{R/L} (t,z)=\pm i \ell \phi'(z) \dot{h}_{R/L}(t,z),
   \label{EOM_h_RL}
\end{eqnarray}
where $\pm$ corresponds to the right-handed/left-handed mode, respectively.
 Using the Fourier mode ${h}_{R/L}(t,z)=H_{R/L}(z)e^{-i\omega t}$, Eq.\eqref{EOM_h_RL} can be reconstructed as the following Schrödinger equation:
\begin{eqnarray}
   \left[ - \frac{d^2 }{dz^2} + V_{\rm{eff}}(z) \right] H_{R/L}(z) = \omega^2 H_{R/L}(z) \ ,
   \label{Scho_eq}
\end{eqnarray}
where we introduced the effective potential  as
\begin{align}
   V_{\rm{eff}}(z) = \pm \ell \omega \phi'(z)
   =\pm \ell  \omega \frac{\eta^2 \sqrt{\frac{\lambda}{2}}}{\cosh^2 \sqrt{\frac{\lambda}{2}}\eta z } \ .
\end{align}
Since the right- and left-handed modes satisfy different equations, circular polarization is expected to be generated as they propagate through the domain wall.
We plotted $V_{\rm{eff}}$ in Fig.\ref{fig_potential}.

\section{Chiral gravitational waves from domain wall}

Now, we are in a position to evaluate the chirality of gravitational waves.
Since the potential vanishes in the asymptotic region, we can express the solution of Eq.(\ref{Scho_eq})  as 
	\begin{eqnarray}
 \label{boundary}
	H_{R/L}(z)&=&
	\begin{cases}
   		 \frac{e^{i\omega z}}{\sqrt{2\omega}}+\mathcal{R}_{R/L}\frac{e^{-i\omega z}}{\sqrt{2\omega}} \qquad(z \to -\infty), \\
         
    		\mathcal{T}_{R/L}\frac{e^{i\omega z}}{\sqrt{2\omega}}  \qquad(z \to \infty),\\
    	 \end{cases}\, \label{BC}
	\end{eqnarray}
where $\mathcal{R}_{R/L}$ and $\mathcal{T}_{R/L}$ are the reflection and transmission coefficients. The point is that the transmission probability $|\mathcal{T}_{R/L}|^2$ depends on the polarizations.

\subsection{Gravitational waves scattered by a domain wall}

In this subsection, we solve the Schrödinger equation \eqref{Scho_eq} to obtain the transmission probability. 

Using the dimensionless variable $x=\omega z$, we can rewrite Eq.\eqref{Scho_eq} as follows:
\begin{eqnarray}
   \left[ - \frac{d^2 }{dx^2} +  \frac{A_{R/L} }{\cosh^2 \alpha x} \right] H_{R/L}(z) =  H_{R/L}(z),
\end{eqnarray}
where we defined the dimensionless parameters;
\begin{eqnarray}
    \alpha= \sqrt{\frac{\lambda}{2}}\frac{\eta}{\omega}
    \ , \qquad
    A_{R/L}=\pm \ell \eta \alpha  \ .
\end{eqnarray}
In the following, the R/L subscripts are omitted. 

Using a new variable $\xi= \tanh \alpha x$ and the
notation $\epsilon = -i/\alpha$ and $s(s+1)=-A/\alpha^2$,
we obtain as
\begin{eqnarray}
 \frac{d }{d\xi} \left[ (1-\xi^2)\frac{d H(\xi)}{d\xi} \right]
 + \left[ s(s+1) - \frac{\epsilon^2 }{1-\xi^2}\right] H(\xi) =  H(\xi).
\end{eqnarray}
After making the transformation $H=(1-\xi)^{\epsilon/2} w$
and $u=(1-\xi)/2$, we have a hypergeometric equation
\begin{eqnarray}
u(1-u)w'' + (\epsilon +1) (1-2u)w' -(\epsilon -s)(\epsilon+s+1) w=0.
\end{eqnarray}
The solution regular at $\xi =1 \,(x\rightarrow \infty)$ is given by
\begin{eqnarray}
H(\xi)= \left( 1- \xi^2\right)^{\frac{\epsilon}{2}}
   F\left[ \epsilon -s , \epsilon + s +1 , \epsilon +1 , \frac{1}{2} \left(1-\xi \right)\right].
\end{eqnarray}
For $x\rightarrow \infty$, we get $\xi \sim 1-2 e^{-2\alpha x}$.
Hence, $(1-\xi^2)^{\epsilon/2} \sim 2^\epsilon e^{i x}$.
Therefore, we obtain
\begin{eqnarray}
H(\xi)\simeq  2^{\epsilon} e^{ix} \qquad (x\to \infty).
\end{eqnarray}
This matches the boundary conditions (\ref{BC}).
For $x\rightarrow -\infty$, we get $\xi \sim -1+2 e^{2\alpha x}$.
Using the following formula
\begin{eqnarray}
 && \hskip -1.2cm F\left[ \alpha, \beta , \gamma , z\right]
 = \frac{\Gamma(\gamma)\Gamma(\gamma  -\alpha-\beta)}{\Gamma(\gamma -\alpha)\Gamma(\gamma-\beta)}
 F\left[ \alpha, \beta , \alpha+\beta +1 -\gamma ,1- z\right]
 \nonumber\\
 && \hskip -1cm + \frac{\Gamma(\gamma)\Gamma(\alpha+\beta -\gamma)}{\Gamma(\alpha)\Gamma(\beta)}(1-z)^{\gamma-\alpha-\beta}
 F\left[ \gamma -\alpha, \gamma - \beta , \gamma +1 -\alpha-\beta ,1- z\right],  \quad
\end{eqnarray}
we can deduce the asymptotic form
\begin{eqnarray}
H(\xi) &=& \frac{\Gamma(\epsilon+1)\Gamma(-\epsilon)}{\Gamma(1+s)\Gamma(-s)}\left( 1- \xi^2\right)^{\frac{\epsilon}{2}}
   F\left[ \epsilon -s , \epsilon + s +1 , \epsilon +1 , \frac{1}{2} \left(1+\xi \right)\right]
   \nonumber\\
 &&  +\frac{\Gamma(\epsilon+1)\Gamma(\epsilon)}{\Gamma(\epsilon -s)\Gamma(\epsilon+s+1)}\left( 1- \xi^2\right)^{\frac{\epsilon}{2}}
 \left(\frac{1+\xi}{2}\right)^{-\epsilon}
   F\left[ 1+s , -s , 1-\epsilon , \frac{1}{2} \left(1+\xi \right)\right]
   \nonumber\\
&\simeq&  \frac{\Gamma(\epsilon+1)\Gamma(-\epsilon)}{\Gamma(1+s)\Gamma(-s)}
2^{\epsilon} e^{-ix}
+ \frac{\Gamma(\epsilon+1)\Gamma(\epsilon)}{\Gamma(\epsilon -s)\Gamma(\epsilon+s+1)} 
2^{\epsilon} e^{ix} \quad (x\to -\infty).
\end{eqnarray}
Comparing the above expression with the boundary condition (\ref{BC}), we obtain the transmission probability as
\begin{eqnarray}
\left|\mathcal{T}\right|^2  = \left|\frac{\Gamma(\epsilon -s)\Gamma(\epsilon+s+1)} {\Gamma(\epsilon+1)\Gamma(\epsilon)}\right|^2.
\label{trans}
\end{eqnarray}

\subsection{Circular polarizations}
In this subsection, we evaluate the transmission probability \eqref{trans} for circular polarization modes of the right-handed and left-handed, separately. 

It is easy to see that
\begin{eqnarray}
\left| \Gamma(\epsilon+1)\Gamma(\epsilon)\right| = \frac{\pi}{\sinh \frac{\pi}{\alpha}} \ .
\end{eqnarray}
The explici form of $s$ is given by
\begin{eqnarray}
    s= \frac{-1\pm \sqrt{1-\frac{4A_{R/L}}{\alpha^2}}}{2}.
\end{eqnarray}
For a left-handed mode, $A_L<0$. Hence $s$ is real. 
In other words, there is no potential barrier.
Hence,  we can deduce the following 
\begin{eqnarray}
\left| \Gamma(\epsilon-s)\Gamma(\epsilon+s+1)\right|^2
&=& \left| \Gamma(i\frac{1}{\alpha}-s)\Gamma(1-i\frac{1}
{\alpha}+s)\right|^2  \nonumber\\
&=&  \left| \frac{\pi}{\sin \pi \left(s-i/\alpha\right)}\right|^2
\nonumber\\
&=&  \frac{\pi^2}{\sin^2 \pi s \cosh^2 \pi/\alpha
+ \cos^2 \pi s \sinh^2 \pi/\alpha }.
\end{eqnarray}
Finally, we obtain the transmission probability
\begin{eqnarray}
\left|\mathcal{T}_L\right|^2  = \frac{\sinh^2 \pi/\alpha}{
 \sinh^2 \pi/\alpha + \sin^2 \pi s  }
 =\frac{\sinh^2 \pi/\alpha}{
 \sinh^2 \pi/\alpha + \cos^2 \left(\frac{\pi}{2}\sqrt{1+4\ell \omega \sqrt{\frac{2}{\lambda}}}\right) } \ .\label{TL}
\end{eqnarray}

On the other hand, for a right-handed mode,  $A_R>0$. 
In this case, depending on the frequency, there are two cases.
For $4A_R<\alpha^2$, $s$ is real. 
In that case, we have the following
\begin{eqnarray}
\left|\mathcal{T}_R\right|^2  = \frac{\sinh^2 \pi/\alpha}{
 \sinh^2 \pi/\alpha + \cos^2 \left(\frac{\pi}{2}  \sqrt{1- 4\ell \omega \sqrt{\frac{2}{\lambda}}} \right) }
 \ . \label{TR1}
\end{eqnarray}
Remarkably, for a specific frequency that satisfies the relation 
\begin{align}
4\ell \omega \sqrt{\frac{2}{\lambda}}=- 4n (n+1)
\end{align}
with integers $n$, there is no reflection. 
For the other case $4A_R>\alpha^2$, $s$ is complex. So, we have $s=-1/2 \pm i \beta/2$ with
$\beta = \sqrt{4 A_R /\alpha^2 -1}$.
Hence, we obtain the transmission probability
\begin{eqnarray}
\left|\mathcal{T}_R\right|^2  = \frac{\sinh^2 \pi/\alpha}{
 \sinh^2 \pi/\alpha + \cosh^2 \left(\frac{\pi}{2} \sqrt{4\ell \omega \sqrt{\frac{2}{\lambda}} -1} \right)}
 \ . \label{TR2}
\end{eqnarray}




\begin{figure}[t]
        \centering
        \includegraphics[width=12cm]{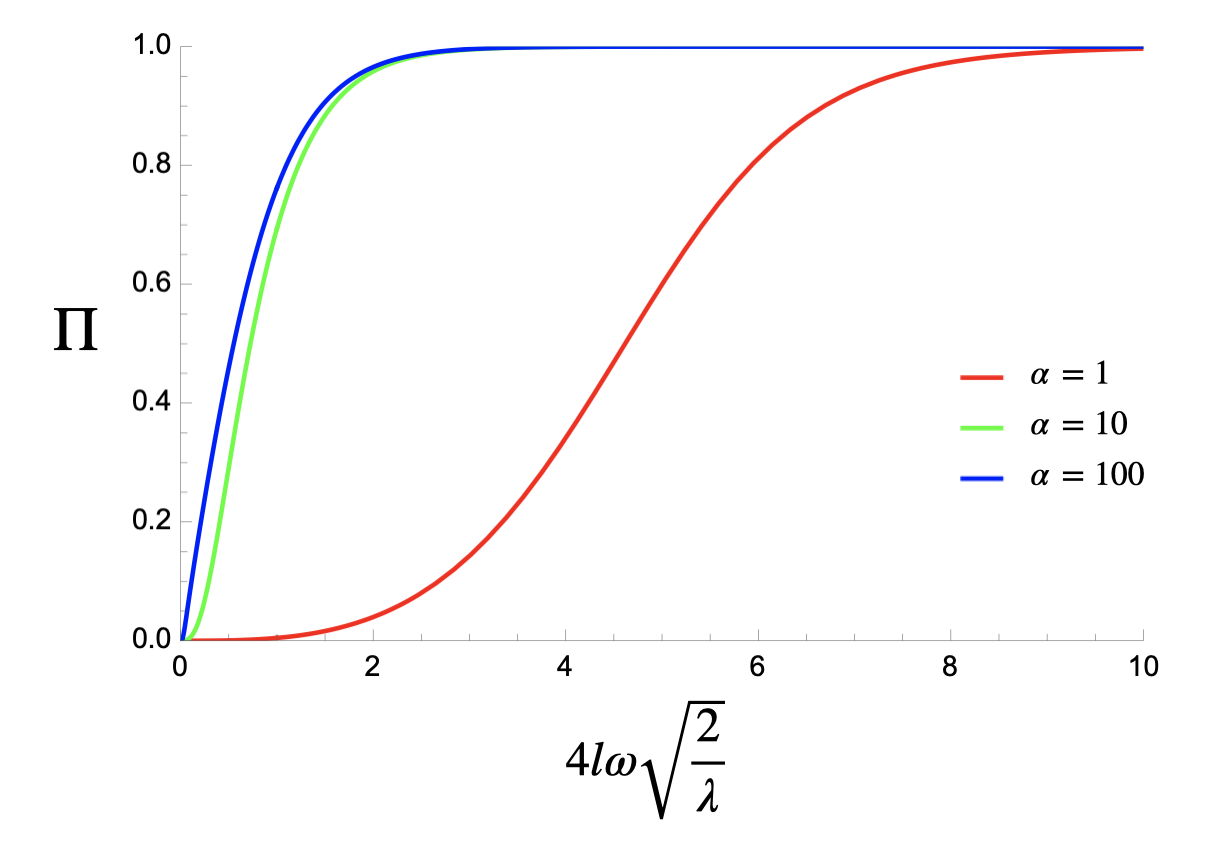}
     \vspace{-6pt}
        \caption{The plot of the degree of circular polarizations.
        We see that the degree of polarizations does not change much even if we change $\alpha$ from 10 to 100.}
      \label{result}
        \vspace{0.7cm}
        \label{Fig_pol}
\end{figure}

The degree of circular polarization can be defined by
\begin{eqnarray}
	\Pi \equiv \frac{|\mathcal{T}_L|^2-|\mathcal{T}_R|^2}{|\mathcal{T}_L|^2+|\mathcal{T}_R|^2}  
    \ .
\end{eqnarray}
Given the analytic formulas (\ref{TL}), (\ref{TR1}),and (\ref{TR2}), we can deduce the degree of circular polarization in each parameter region.
In the case $4 A_R <\alpha^2$, we have the formula
\begin{eqnarray}
	\Pi 
    = \frac{ \cos^2 \left(\frac{\pi}{2}  \sqrt{1- 4\ell \omega \sqrt{\frac{2}{\lambda}}}\right)- \cos^2 \left(\frac{\pi}{2}  \sqrt{1+ 4\ell \omega \sqrt{\frac{2}{\lambda}}} \right)}{
 2\sinh^2 \pi/\alpha +  \cos^2 \left(\frac{\pi}{2}  \sqrt{1- 4\ell \omega \sqrt{\frac{2}{\lambda}}} \right) + \cos^2 \left(\frac{\pi}{2}  \sqrt{1 + 4\ell \omega \sqrt{\frac{2}{\lambda}}} \right) }
    \ .
\end{eqnarray}
In the case $4 A_R>\alpha^2$,
the degree of circular polarization reads
\begin{eqnarray}
	\Pi 
    = \frac{\cosh^2 \left(\frac{\pi}{2} \sqrt{4\ell \omega \sqrt{\frac{2}{\lambda}} -1} \right)- \cos^2 \left(\frac{\pi}{2}  \sqrt{1+ 4\ell \omega \sqrt{\frac{2}{\lambda}}} \right)}{
 2\sinh^2 \pi/\alpha + \cosh^2 \left(\frac{\pi}{2} \sqrt{4\ell \omega \sqrt{\frac{2}{\lambda}} -1} \right) + \cos^2 \left(\frac{\pi}{2}  \sqrt{1 + 4\ell \omega \sqrt{\frac{2}{\lambda}}} \right) }
    \ .
\end{eqnarray}
Thus, we have succeeded in obtaining analytic formulas for the degree of circular polarizations.

Since CMB observations give a constraint on the tension of domain walls $\sigma < (0.93\ {\rm{MeV}})^3 $~\cite{Lazanu:2015fua}, we asuume $\eta \sim {\rm{MeV}}$.
Hence, the parameter $\alpha$ is typically large.
In Fig.\ref{Fig_pol}, we plotted $\Pi$ for specific parameters.
As can be seen in Fig.\ref{Fig_pol}, no differences can be observed from the behavior of $\Pi$ for large values of $\alpha$.
Thus, the symmetry breaking scale does not affect the behavior of the degree of circular polarization.
On the other hand, from Fig.\ref{Fig_pol}, we see that $\Pi$ is close to 1 for $4 A_R \gg \alpha^2$.
Thus, the condition for obtaining chiral gravitational waves after transmission through domain walls can be written as
\begin{eqnarray}
    \omega \gg \frac{1}{4\ell} \sqrt{\frac{\lambda}{2}}  \ .
\end{eqnarray}
 Apparently, the condition is completely independent of the energy scale of the symmetry breakdown. 
 Hence, the generation of circular polarization occurs irrespective of the tension of the domain wall.

\section{Conclusion}
Parity violation in gravity is an intriguing possibility. 
In the context of Einstein gravity, the Chern-Simons term can be considered. However, it suffers from the ghost instability.
Interestingly, in teleparallel gravity, the healthy parity violating term, the so-called Nieh-Yan term, is known.
We studied the scattering of gravitational waves by axion domain walls in teleparallel gravity with the Nieh-Yan term. 
Since a domain wall causes the parity violation, the transmitted 
gravitational waves also exhibited the parity violation.
We calculated the degree of circular polarization of gravitational waves. 
We have shown that chiral gravitational waves are ubiquitous.
It turned out that gravitational waves with wavelength shorter than the characteristic length scale $\ell$  will be circularly polarized after the transmission of domain walls.
Remarkably, we found that the condition for obtaining chiral gravitational waves after passing through domain walls does not depend on the tension of a domain wall. Thus, this happens for any scenario as long as axions exist.
In particular, any scattering of gravitational waves with axion domain walls can be a source of the chiral gravitational waves. 
Therefore, the chirality of gravitational waves could be a probe of axions. 

Although we have considered the scattering of gravitational waves by domain walls, we can repeat the same calculations for other situations such as the axion clump~\cite{Marsh:2015xka}. Since there always exists birefringence, we would have the circular polarization. 

In contrast to the Chern-Simons term, the Nieh-Yan term does not cause ghost instability. Hence, it is possible to study its role in black hole physics and cosmology.
It is interesting to investigate quasi-normal modes in Nieh-Yan gravity following the analysis in \cite{Kanno:2025deb}.
We can also study parametric amplification of gravitational waves~\cite{Jung:2020aem,Chu:2020iil}.
We leave these works for the future.

\section*{Acknowledgments}

J.\ S. was in part supported by JSPS KAKENHI Grant Numbers JP23K22491, JP24K21548, JP25H02186.
M.T. was in part supported by the YAMAGUCHI UNIVERSITY FUND and the Sumitomo Foundation under Grant for Basic Science Research (Grant No. 2502479).

\bibliography{references}
\bibliographystyle{unsrt}
\end{document}